\documentclass[11pt,a4paper]{article}
\usepackage{amsmath,amssymb}
\usepackage{epsfig,graphicx}
\usepackage{cite}
\usepackage{amsbsy}

\begin{document}

\title{\bf  Propagators of charged particles in an external magnetic field, expanded
over Landau levels} 

\author{A.~V.~Kuznetsov\footnote{{\bf e-mail}: avkuzn@uniyar.ac.ru},
A~A.~Okrugin\footnote{{\bf e-mail}: alexander.okrugin@gmail.com},
A.~M.~Shitova\footnote{{\bf e-mail}: pick@mail.ru}
\\
\small{\em Division of Theoretical Physics, Department of Physics,} \\
\small{\em Yaroslavl State P.G.~Demidov University} \\
\small{\em Sovietskaya 14, 150000 Yaroslavl, Russian Federation }
}

\date{}

\maketitle

\begin{abstract}
 Various forms of expressions for the propagators of charged
particles in a constant magnetic field that should be used for investigations of
electroweak processes in external uniform magnetic field are discussed. Formulas
for the propagators of the Standard Model charged $W$- and scalar $\Phi$-bosons in
an arbitrary $\xi$-gauge, expanded over Landau levels, are derived for the first
time. \\

Keywords; propagator; external magnetic field; Landau levels; $\xi$-gauge;
self-energy operator.\\

\end{abstract}

\vfill

\begin{center}
Preprint of an article submitted for consideration 

in International Journal of Modern Physics A 

\copyright 2015 World Scientific Publishing Company,

http://www.worldscientific.com/worldscinet/ijmpa
\end{center}

\newpage

%%%%%%%%%%%%%%%%%%%%%%%%%%%%%%%%%%%%%%%%%%%%%%%%%%%%%%%%%
\section{Introduction} \label{sec:Introduction}
%%%%%%%%%%%%%%%%%%%%%%%%%%%%%%%%%%%%%%%%%%%%%%%%%%%%%%%%%

Recently, the study of electroweak processes in ultra-strong magnetic fields
became extremely actual. There are several cases when such fields could be
possible.

The conception of magnetars\cite{Duncan:1992} (a kind of neutron stars with extremely
high magnetic fields) encouraged the
scientists to reanalyze the electroweak processes in external magnetized 
media with large scale magnetic fields. According to the McGill Online Magnetar
Catalog \cite{McGillOnline}, there are 28 objects expected to be magnetars. 
Approximately a half of them are the soft gamma-ray repeaters (SGR) 
and the others are the anomalous X-ray pulsars (AXP). 
The above-mentioned magnetic field scales are greater than the so-called Schwinger field limit---a natural scale for
the magnetic field strength:\footnote{ We use the Planck units:
$\hbar = 1, c = 1$, and the Minkowski metric with the signature $(+{} -{} -{} -)$.}
$B_e = m_e^2/e \simeq 4.41 \times 10^{13}$\,G. Even stronger magnetic fields could
exist in the early Universe. Note that for the electric field instead of the
magnetic field the shown limit is crucial because of the intensive vacuum
electron-positron pair creation that is equal to the short circuit of an electric field generator. 
On the other hand, the vacuum is stable in a magnetic field, so its strength could exceed the critical value $B_e$. 
Moreover, the magnetic field could play a stabilizing role in the case when it is perpendicular
to the electric one. In such a configuration, the electric field could be even greater than $B_e$.
The invariant form of the stability condition is as follows:
$$F^{\mu \nu} F_{\mu \nu} = 2\left(B^2 - {\cal E}^2\right) \geqslant 0\,,$$ \noindent
where $F^{\mu \nu}$ is the electromagnetic field tensor.

Nowadays, in our Universe from stars to galaxy clusters, magnetic fields of
different scales from mG to G are observed. There is an open question about the
causes of observable large scale magnetic fields. On the one hand,
the magnetohydrodynamic mechanism explains successfully the rate of the observed
fields, but for its realizations some seed fields of order $10^{-21}$\,--\,$10^{-19}$\,G
are required \cite{Kunze:2013}. In this case, the question emerges when and how do
the first seed magnetic fields appear? 
In the early Universe, in the interval between the stages of the QCD phase transition and nucleosynthesis, 
very strong magnetic fields, the so called ``primary'' fields, 
in principle, could exist, with an initial strength of the order
$10^{23}$~G \cite{Vachaspati:1991}
and even more ($\sim 10^{33}$~G) \cite{Ambjorn:1992}. Their evolution
during the expansion of the Universe
could determine the existence at the present stage of coherent
large-scale ($\sim$ 100~kpc) 
magnetic fields with an intensity $\sim 10^{-21}$~G. 
These fields, in turn, could be enhanced by
the galactic dynamo mechanism to the observed values of galactic 
magnetic fields $\sim 10^{-6}$~G.
Possible origins of primary strong magnetic fields and dynamics of their
evolution in the expanding Universe are the subject of intense
research (see, for example, the surveys~\cite{Grasso:2001,Giovannini:2004,Kandus:2011}
and references cited therein).

Finally, one more case when strong magnetic fields could be possibly observed
are the contemporary elementary particle accelerators. The fact of observing of the fields
of magnetar scales in the conditions of the Earth laboratories seems to be suprising
but the fields of such magnitudes could really be formed during non-central
heavy-ion collisions \cite{Skokov:2009}.

While solving the series of principal problems of the interactions of charged 
particles with an electromagnetic field, the method has gained a great value, in
which the impact of external fields is considered on the base of the direct
solutions of wave equations in the external electromagnetic field instead of using the
perturbation theory. The method was first proposed by Sokolov and later
developed by Sokolov and Ternov while solving the problem of the synchrotron
radiation. The corresponding idea is now known as the Furry picture. The calculation
of the specific physical phenomena supposes the using of the Feynman diagram
technique with the following generalization: in the initial and final states, a charged
particle is in the external field and is described by the solution of the wave
equation in this field, and the internal lines of the charged particles
correspond to the propagators built on a base of that solution. The detailed description
of the calculation technique of the processes in external fields one can find,
for example, in reviews and book, see e.g. Refs.
\cite{Ritus:1979,Itzykson:1980,Papanian:1986,Shabad:1988,Ternov:1989,Kuznetsov:2003}.
The method is very useful in the case
of strong magnetic fields when the account of the field influence can not be made
with the help of the perturbation theory. By virtue of the vacuum stability in
superstrong magnetic fields, it is possible to study processes in the fields with
the intensities significantly above the critical value $B_e$.

The extent of influence of the external field on the propagation of a charged particle
is determined by its specific charge i.e. the ratio of the particle charge to its mass.
So, the light charged fermions especially electrons are mainly sensitive to the external field influence.

The expression for the exact electron propagator in the constant uniform magnetic
field was first obtained by J. Schwinger\cite{Schwinger:1951}
in the Fock-Schwinger\cite{Fock:1937} proper time formalism. There exists a number
of works where another forms of propagators were derived. For example
the case of a superstrong magnetic field was analyzed in Ref.~\cite{Loskutov:1976} 
and the contribution from the ground Landau level to the electron propagator was obtained. 
In Ref.~\cite{Chodos:1990}, the propagator was transformed from
the Schwinger form\cite{Schwinger:1951} to the expansion over all Landau levels.
In Ref.~\cite{Chyi:2000}, the expansion of the electron propagator in the
power series of the intensity of a magnetic field was presented. The exact propagator of an
electron in a constant uniform magnetic field as the sum over Landau levels was
obtained in Ref.~\cite{Kuznetsov:2011_Okr} from exact solutions of the Dirac equation 
by a direct derivation of the standard methods of the quantum field theory.

In our opinion, a knowledge of different representations of the charged particle propagators 
in an external magnetic field and the conditions of their applicability would be very
important. There exist several precedents when misunderstanding of such conditions
led to incorrect studies. For instance, a calculation of the neutrino self-energy operator in a magnetic field 
was performed in Refs.~\cite{Elizalde:2002,Elizalde:2004} by analyzing the one-loop diagram
$\nu \to e^-\, W^+ \to \nu$. The authors restricted themselves
by the contribution to the electron propagator from the ground Landau level.
As there was shown in Ref.~\cite{Kuznetsov:2006}, in that case the contribution 
from the ground Landau level did not dominate due to the large electron virtuality, 
and contributions from other levels were of the same order. Ignoring
such a fact led the authors\cite{Elizalde:2002,Elizalde:2004} to incorrect results. 
Another example of this kind was an attempt to reanalyze
the probability of the neutrino decay $\nu \to e^- W^+$ in an external magnetic field
in the limit of ultra-high neutrino energies, calculated via the imaginary part
of the one-loop amplitude of the transition $\nu \to e^- \, W^+ \to \nu$.
Initially, the result was obtained in Ref.~\cite{Erdas:2003}, later the other
authors\cite{Bhattacharya:2009} repeated the calculation and insisted on another
result. The third independent calculation\cite{Kuznetsov:2010_PLB}
confirmed the result of Ref.~\cite{Erdas:2003}. The most likely cause of the
error in Ref.~\cite{Bhattacharya:2009} was that the authors restricted themselves by
only linear terms in the expansion of the $W$-boson propagator over
the electromagnetic tensor $F^{\mu\nu}$ whereas the quadratic terms were essential as well.

Among the works dedicated to the exploration of the propagators of charged particles
in the external field one should note Ref.~\cite{Erdas:2000}, where the
calculation of a neutrino self-energy operator in a magnetic field in an arbitrary
$\xi$-gauge was held. It was demonstrated that in spite of the fact that
the self-energy operator depended on the gauge parameter $\xi$, the observed neutrino
properties arising from its dispersion law, as expected, were gauge-invariant.

Analyses of literature showed that there are no methodically important
expressions for propagators of charged $W$- and $\Phi$-bosons as expansions
over Landau levels. In the present work we calculate, for the sake of completeness, various
representations of propagators of charged particles in a constant uniform magnetic
field, that could be useful for the analyses of the electroweak processes in strong magnetic
fields. Different methodically important transitions from one representation to
another are also performed. In Sec.~\ref{sec:PropagatorsFS}, the general forms of the 
electron, charged $W$- and $\Phi$-boson propagators in the Fock-Schwinger proper
time formalism are presented. In Sec.~\ref{sec:phase}, several important
comments on the translation and gauge noninvariant phases are made. Further in
Sec.~\ref{sec:Propagatorsweak}, expansions are presented of propagators in the
weak-field approximation. In Sec.~\ref{sec:ElectronPropagatorsLandau}, the procedure of
transformation of the electron propagator to the form where it is presented as the sum
over all Landau levels is shown. The formulas for propagators of the charged $W$-
and scalar charged $\Phi$-bosons in an arbitrary $\xi$-gauge expanded over Landau levels
similarly to the electron propagator expansion are derived in Sec.~\ref{sec:PropagatorsLandau} for the first
time. Some details of calculations are presented in the appendices.

We use the notation for the 4-vectors $X^\mu = (t, x, y, z)$; the index numbers
designate the different 4-vectors.  It is implied in the paper that the squared 
masses have infinitely small imaginary parts
$m^2 \to m^2 - \mathrm{i} \epsilon$.

%%%%%%%%%%%%%%%%%%%%%%%%%%%%%%%%%%%%%%%%%%%%%%%%%%%%%%%%%
\section{Propagators in the Fock-Schwinger proper time formalism}
\label{sec:PropagatorsFS}
%%%%%%%%%%%%%%%%%%%%%%%%%%%%%%%%%%%%%%%%%%%%%%%%%%%%%%%%%

The electron propagator in a constant uniform magnetic field in the Fock-Schwinger proper
time formalism can be presented in the form
\begin{equation}
S^{(e)} (X_1,X_2) = \mathrm{e}^{\mbox{\normalsize $\mathrm{i} \Phi (X_1,X_2)$}} S (X_1-X_2)\,.
\label{eq:S0}
\end{equation}

\noindent Here, $S (X)$ is the translation and gauge invariant part of the propagator:
\begin{gather}
S (X)=-\frac{\mathrm{i} \beta}{2 (4 \pi)^2} \,
\int\limits_0^\infty \frac{\mathrm{d} s}{s \sin (\beta s)}
 \bigg \lbrace \frac{1}{s} \bigg [
\cos (\beta s)
(X \tilde \varphi \tilde \varphi \gamma) -\mathrm{i} \sin
(\beta s) (X \tilde \varphi \gamma) \gamma_5 \bigg ] \nonumber\\
-\frac{\beta}{\sin (\beta s)} (X \varphi
\varphi \gamma) + m_e \bigg[2 \cos (\beta s)
 -\sin (\beta s) (\gamma \varphi \gamma)
\bigg] \bigg \rbrace \nonumber\\ \times \exp \bigg \{
- \mathrm{i} \bigg [ m_e^2 s + \frac{(X \tilde \varphi \tilde
\varphi X)} {4 s} -\frac{\beta}{4 \tan (\beta s)} (X \varphi \varphi X) \bigg] \bigg \}\,, \label{eq:SB}
\end{gather}
\noindent where $\beta = e B$, $e$ is the elementary charge, $m_e$ is the electron
mass, $X_\mu = (X_1-X_2)_\mu$, $\varphi_{\alpha \beta} = F_{\alpha \beta} /B$
is the dimensionless electromagnetic tensor, ${\tilde \varphi}_{\alpha \beta} =
\frac{1}{2} \, \varepsilon_{\alpha \beta \mu \nu} \, \varphi^{\mu \nu}$ is the dual
dimensionless tensor ($\varepsilon^{0123} = - \varepsilon_{0123} = +1$); tensor
indexes for 4-vectors and tensors in parentheses are assumed to be contracted consecutively, 
for instance: $(X \varphi \varphi \gamma) =
X^{\alpha} \varphi_{\alpha \beta} \varphi^{\beta \mu} \gamma_{\mu}$.

Integration with respect to the variable $s$ (corresponding to the proper time) in
the expression (\ref{eq:SB}) needs to be redefined due to the poles of the integrand
in the points $s = \pi k/\beta$, where $k = 0, 1, 2\dots$ It is assumed that
the integration is performed in a complex plane $s$ over the contour that begins
in the point $s = 0$ and lies below the real axis closely to it. This contour may also
be rotated to the negative imaginary semiaxis (see below).

The phase $\Phi (X_1,X_2)$ in the formula~(\ref{eq:S0}) is translationally and gauge
noninvariant and can be defined in terms of an integral along an arbitrary contour as:
\begin{eqnarray}
\Phi (X_1,X_2)=- \, e \int\limits^{\mbox{\normalsize
$X_2$}}_{\mbox{\normalsize $X_1$}} \mathrm{d} X_\mu \, K^\mu (X)\,,
\label{eq:FKA}\\
K^\mu (X) = A^\mu (X) + \frac{1}{2} F^{\mu \nu} (X - X_2)_\nu\,.
\label{eq:KA}
\end{eqnarray}

Similarly to the equation~(\ref{eq:S0}), one can define the propagators of the $W$-boson
and the charged scalar $\Phi$-boson in a magnetic field (note that we assume
negatively charged $W^-$ and $\Phi^-$-bosons to be particles):
\begin{eqnarray}
G^{(W)}_{\mu \nu} (X_1,X_2) = \mathrm{e}^{\mbox{\normalsize
$\mathrm{i} \Phi (X_1,X_2)$}} \, G_{\mu \nu} (X_1-X_2)\,,
\label{eq:G0}\\
D^{(\Phi)} (X_1,X_2)=\mathrm{e}^{\mbox{\normalsize $\mathrm{i}
\Phi (X_1,X_2)$}} \, D (X_1-X_2)\,, \label{eq:D0}
\end{eqnarray}
\noindent where the phase $\Phi (X_1,X_2)$ is defined by the same
formulas~(\ref{eq:FKA}), (\ref{eq:KA}).

It is often convenient to use the Fourier transforms of the translationally
invariant parts of propagators:
\begin{eqnarray}
S (X) &=& \int \frac{\mathrm{d}^4 q}{(2 \pi)^4} \, S (q) \,
\mathrm{e}^{- \mathrm{i} q X}\,,
\label{eq:S_Four}\\
&&\nonumber\\
G_{\mu \nu} (X) &=& \int \frac{\mathrm{d}^4 q}{(2 \pi)^4} \,
G_{\mu \nu} (q)\,\mathrm{e}^{- \mathrm{i} q X}\,,
\label{eq:G_Four}\\
&&\nonumber\\
D (X) &=& \int \frac{\mathrm{d}^4 q}{(2 \pi)^4} \, D
(q)\,\mathrm{e}^{- \mathrm{i} q X}\,. \label{eq:D_Four}
\end{eqnarray}
\noindent One can obtain the Fourier transform
of the electron propagator from Eqs. (\ref{eq:SB}) and (\ref{eq:S_Four}) in the form:
\begin{eqnarray}
S (q)=\int\limits_0^{\infty}\!\! \frac{\mathrm{d} s}{\cos (\beta
s)}\,
\exp\bigg[- \mathrm{i} s \bigg(m_e^2- q_{\|}^2
 + q_{\perp}^2\, \frac{\tan (\beta s)}{\beta s} \bigg) \bigg] && \nonumber\\ \times
\biggl\{\left[(q \gamma)_{\|} + m_e \right] \bigg[ \cos (\beta s)
- \,\frac{ (\gamma \varphi \gamma)}{2} \, \sin (\beta s)
\bigg]-\frac{(q \gamma)_{\perp}}{\cos (\beta s)}\biggr\}\,,\, &&
\label{eq:S(q)}
\end{eqnarray}
\noindent where the following denotations for multiplications of longitudinal
and transverse 4-vectors are used $(q \gamma)_{\|} = (q \, \tilde \varphi
\tilde \varphi \,\gamma)$, $(q \gamma)_{\perp} = (q \,\varphi \varphi \, \gamma)$.
In the coordinate system where the 3d space axis is directed along the vector of an
external magnetic field $\vec{B}$, the 4-vectors with the indexes $\perp$ and $\|$
belong to the Euclidean $\{1, 2\}$-subspace and the Minkowski $\{0,3\}$-subspace
respectively. For instance, $p_{\perp} = (0,p_1,p_2,0)$, $p_{\|} =
(p_0,0,0,p_3)$.

The Fourier transforms of the propagators of the $W$-boson (\ref{eq:G0}),
(\ref{eq:G_Four}) and the charged scalar $\Phi$-boson
(\ref{eq:D0}), (\ref{eq:D_Four}) depend on the gauge selection.
In an arbitrary $\xi$-gauge they take the form \cite{Erdas:2000}:

\begin{gather} G_{\mu \nu} (q) = - \int\limits_0^{\infty}
\frac{\mathrm{d} s}{\cos (\beta s)} \mathrm{e}^{\mathrm{i} s
\left(q_{\|}^2 - q_{\perp}^2 \, \tan (\beta s)/(\beta s) \right)}
 \biggl\{ \mathrm{e}^{- \mathrm{i} s m_W^2} \biggl[
g_{\mu \nu} + (\varphi \varphi)_{\mu \nu} \left( 1 - \cos (2 \beta
s) \right)
 \nonumber\\
- \varphi_{\mu \nu} \, \sin (2 \beta s) \biggr]
- \frac{1}{m_W^2}\biggl[ \biggl( q_\mu + (\varphi q)_\mu \, \tan
(\beta s) \biggr) \biggl( q_\nu + (q
\varphi)_\nu \, \tan (\beta s) \biggr)
\nonumber\\
 + \mathrm{i} \,
\frac{\beta}{2} \, \biggl( \varphi_{\mu
\nu} - (\varphi \varphi)_{\mu \nu} \, \tan (\beta s)\biggr) \biggr]
 \left( \mathrm{e}^{- \mathrm{i} s m_W^2} - \mathrm{e}^{- \mathrm{i} s \,\xi \,
m_W^2}\right) \biggr\}\,, 
\label{eq:G(q)xi}\end{gather}
\vspace{-4pt}
\begin{equation}
D (q)=\int\limits_0^{\infty} \frac{\mathrm{d} s}{\cos (\beta s)}
\mathrm{e}^{- \mathrm{i} s \left(\xi \, m_W^2- q_{\|}^2 +
q_{\perp}^2 \, \tan (\beta s)/(\beta s) \right)}\,. \label{eq:D(q)xi}
\end{equation}
In the Feynman gauge, where $\xi = 1$, the Fourier transform of the $W$-boson propagator
significantly simplifies \cite{Erdas:1990}:
\begin{gather}
 G_{\mu \nu} (q) = - \int\limits_0^{\infty} \frac{\mathrm{d}
s}{\cos (\beta s)} \mathrm{e}^{- \mathrm{i} s \left(m_W^2-
q_{\|}^2
 + \tan (\beta s)/(\beta s)\,q_{\perp}^2 \right)}
\nonumber\\\times \biggl[ g_{\mu \nu} + (\varphi
\varphi)_{\mu \nu} \left( 1 - \cos (2 \beta s) \right) -
\varphi_{\mu \nu} \, \sin (2 \beta s) \biggr]\,.
\label{eq:G(q)}
\end{gather}
\noindent The Fourier transform of the propagator of the charged scalar
$\Phi$-boson in the Feynman gauge has the form:
\begin{equation}
D (q) = \int\limits_0^{\infty} \frac{\mathrm{d} s}{\cos (\beta s)}
\exp\left[- \mathrm{i} s \left(m_W^2- q_{\|}^2 + \frac{\tan (\beta
s)}{\beta s}\,q_{\perp}^2 \right) \right]\,.
\label{eq:D(q)}
\end{equation}

%%%%%%%%%%%%%%%%%%%%%%%%%%%%%%%%%%%%%%%%%%%%%%%%%%%%%%%%%
\section{Comment on the noninvariant phase}
\label{sec:phase}
%%%%%%%%%%%%%%%%%%%%%%%%%%%%%%%%%%%%%%%%%%%%%%%%%%%%%%%%%
 
The expressions~(\ref{eq:FKA}), (\ref{eq:KA}) for translationally and gauge noninvariant phase $\Phi (X_1,X_2)$
written in a covariant form look rather cumbersome. Some authors prefer to write down the phase in
more compact form, fixing the gauge by choosing 4-potential as $A^\mu (X) = (0, 0, x\, B, 0)$, to obtain:
\begin{equation}
\Phi (X,X^{\,\prime}) = -\,\frac{e B}{2} \, (x + x^{\,\prime})(y - y^{\,\prime})\,.
\label{eq:F_gauge_fixed}
\end{equation}

Nevertheless, the exactly covariant form of the phase~(\ref{eq:FKA}), (\ref{eq:KA})
appears to be much more efficient in an analysis of the closed loops containing several
propagators of charged particles. Note that by virtue of the property of the
4-vector~(\ref{eq:KA}): $\partial_\mu K_\nu - \partial_\nu K_\mu = 0$, the path
of integration from $X_1$ to $X_2$ in Eq.~(\ref{eq:FKA}) is arbitrary. In the case
of the two-vertex loop, the sum of the phases contributed to the amplitude is zero:

\begin{equation}
\Phi (X_1,X_2) + \Phi (X_2,X_1) = 0\,. \label{eq:LF}
\end{equation}

In the presence of the three or more vertices, there appears a nonzero total phase from
all propagators, which is nonetheless translationally and gauge invariant. This
can be easily shown by rewriting the 4-potential of the constant uniform
external field in an arbitrary gauge in the following form:

\begin{equation}
A^\mu (X) = \frac{1}{2} \, X_\nu F^{\nu \mu} + \partial^\mu \chi
(X)\,, \label{eq:A(x)}
\end{equation}
where $\chi (X)$ is an arbitrary function. Given Eq. (\ref{eq:A(x)}), one 
automatically obtains that $\partial^\mu A^\nu - \partial^\nu A^\mu =
F^{\mu \nu}$. Integrating~(\ref{eq:FKA}) with taking into account the
relation~(\ref{eq:A(x)}), we obtain
\begin{equation}
\Phi(X_1,X_2) = - \frac{e}{2} \, (X_1 F X_2) - e \, [\chi (X_2) - \chi (X_1)]\,.
\label{eq:phase}
\end{equation}
It is seen from Eq.~(\ref{eq:phase}) that during the summation of phases
in a closed loop, the terms containing the function $\chi$ is completely cancelled, 
providing the gauge invariance. It can be easily checked that the sum of the 
expressions~(\ref{eq:phase}) in a closed loop would be translationally
invariant also.

For instance, for the three and four propagators in a loop, the sum of phases takes
the form:
\begin{eqnarray}
\Phi(X_1, X_2) + \Phi(X_2, X_3) + \Phi(X_3, X_1) =
 -\frac{e}{2} \, (X_1 - X_2)_\mu F^{\mu \nu} (X_2 - X_3)_\nu\,,
 \label{eq:phase3}
\end{eqnarray}

\begin{eqnarray} \Phi(X_1, X_2)+\Phi(X_2, X_3)+\Phi(X_3,
X_4)+\Phi(X_4, X_1) \nonumber\\=- \frac{e}{2} (X_1 - X_3)_\mu
F^{\mu \nu} (X_2 - X_4)_\nu\,.\label{eq:phase4}
\end{eqnarray}

In a general case, the sum of $n$ phases can be presented in the following form:

\begin{eqnarray}
\Phi_{tot} = - \frac{e}{2} \sum \limits_{i=1}^n (X_i F X_{i+1})
{\bigg |}_{X_{n+1} \equiv X_1}
 = - \frac{e}{2} \sum \limits_{l=2}^{n-1}
\sum \limits_{k=1}^{l-1} (Z_k F Z_l)\,, \label{eq:phase_n}
\end{eqnarray}
where $$Z_i \, = \, X_i - X_{i+1}\,.$$

%%%%%%%%%%%%%%%%%%%%%%%%%%%%%%%%%%%%%%%%%%%%%%%%%%%%%%%%%
\section{Propagators expanded over the weak field}
\label{sec:Propagatorsweak}
%%%%%%%%%%%%%%%%%%%%%%%%%%%%%%%%%%%%%%%%%%%%%%%%%%%%%%%%%

Using the exact expressions~(\ref{eq:S(q)}) and (\ref{eq:G(q)}) makes the
calculations sufficiently cumbersome. At the same time, magnetic fields existing
in Nature, except the early Universe, are always weak compared to the critical field
for the $W$-boson, $B_W = m_W^2/e \simeq 10^{24}$\,G. Hence propagators of the $W$-boson
and the charged scalar $\Phi$-boson could be expanded in the series over the powers
of $\beta = e B$ assuming $\beta$ to be a small parameter. Keeping the terms up to
the second order for the $W$-boson propagator the in Feynman gauge one can get:
\begin{eqnarray}
&& G_{\mu \nu} (q) = - \mathrm{i} \, \frac{g_{\mu \nu}}{q^2 -
m_W^2} \, - \, \beta \, \frac{2 \, \varphi_{\mu \nu}}{(q^2 -
m_W^2)^2} +
\label{eq:G(q)<} \mathrm{i}\, \beta^2 \biggl[g_{\mu \nu} \left(\frac{1}{(q^2 -m_W^2)^3} \right.\nonumber\\&&+\left. \frac{2 \, q_{\perp}^2}{(q^2 - m_W^2)^4} \right) + 4 \, (\varphi \varphi)_{\mu \nu} \,\frac{1}{(q^2 - m_W^2)^3}
\biggr] + {\cal O}(\beta^3)\,.
\end{eqnarray}
\noindent It is easy to perform the similar calculation for the $W$-boson in an arbitrary
$\xi$-gauge, but the final expression is too cumbersome.

Comparing~(\ref{eq:G(q)}) and (\ref{eq:D(q)}) one can see that the $\Phi$-boson
propagator $D (q)$ differs only by the sign from the coefficient before
$g_{\mu \nu}$ in the expansion of the propagator $G_{\mu \nu} (q)$ over the three
independent tensor structures. So,
\begin{eqnarray}
&& D (q) = \frac{\mathrm{i}}{q^2 - m_W^2} - \mathrm{i}\, \beta^2
\bigg(\frac{1}{(q^2
- m_W^2)^3} + \frac{2 \, q_{\perp}^2}{(q^2 -
m_W^2)^4} \bigg)+{\cal O}(\beta^3)\,. \label{eq:D(q)<}
\end{eqnarray}

The asymptotic expression for the electron propagator $S (q)$ in the case when the field
intensity is the minor dimensional parameter of the problem
$\beta \ll m_e^2 \ll m_W^2$ can be obtained in the similar way. In the case of the weak
field approximation, the expansion for the electron propagator can be derived in the
form \cite{Chyi:2000}:
\begin{eqnarray}
&& S (q) = \mathrm{i} \, \frac{(q \gamma) + m_e}{q^2 - m_e^2} \, +
\, \beta \, \frac{(q \gamma)_{\|} + m_e}{2 (q^2 - m_e^2)^2}
\,(\gamma \varphi \gamma) 
\nonumber\\
&& + \beta^2 \, \frac{2 \, \mathrm{i} \left[(q_{\|}^2 - m_e^2) \,
(q \gamma)_{\perp} - q_{\perp}^2 \, ((q \gamma)_{\|} + m_e)
\right]}
{(q^2 - m_e^2)^4} + {\cal O}(\beta^3)\,. \label{eq:S(q)<}
\end{eqnarray}
Note that in this expansion, the contribution from the region of the small virtual
momenta, $q^2 \sim m_e^2 \ll m_W^2$, increases in each next terms. Wherein in
the case of the moderately strong field, $m_e^2 \ll \beta \ll m_W^2$, the
expansion~(\ref{eq:S(q)<}) is inapplicable and it is necessary to use the exact
expression~(\ref{eq:S(q)}) for the propagator.

%%%%%%%%%%%%%%%%%%%%%%%%%%%%%%%%%%%%%%%%%%%%%%%%%%%%%%%%%
\section{Electron propagator as an expansion over the Landau levels}
\label{sec:ElectronPropagatorsLandau}
%%%%%%%%%%%%%%%%%%%%%%%%%%%%%%%%%%%%%%%%%%%%%%%%%%%%%%%%%

Let us rewrite the Fourier transform~(\ref{eq:S(q)}) of the translationally and gauge
invariant part of the propagator introducing a new variable of integration
$v = \beta s$:
\begin{eqnarray}
 S (q)=\frac{1}{\beta} \int\limits_0^{\infty} \mathrm{d} v \,
\exp (- \mathrm{i} \rho v) \, \biggl\{ \left[(q \gamma)_{\|} + m_e
\right] \, f_1 (v)  &&
\nonumber\\
 -\left[(q \gamma)_{\|} + m_e \right] \frac{ (\gamma \varphi
\gamma)}{2} \, f_2 (v) - (q \gamma)_{\perp} \, f_3 (v) \biggr\}
\,, \label{eq:S(q)_2}
\end{eqnarray}

\noindent where we use the notations:

\begin{eqnarray}
f_1 (v) &=& \exp (- \mathrm{i} \, \alpha \, \tan v) \,,
\nonumber\\[2mm]
f_2 (v) &=& \tan v \, \exp (- \mathrm{i} \, \alpha \, \tan v) \,,
\label{eq:S(q)_f_123}\\[2mm]
f_3 (v) &=& \frac{1}{\cos^2 v} \, \exp (- \mathrm{i} \, \alpha \,
\tan v) \,,\nonumber
\end{eqnarray}
\noindent and also $\rho = ( m_e^2 - q_{\|}^2 )/\beta$, $\alpha = q_{\perp}^2 /\beta$.
Because of the periodic character of the functions $f_j (v) \, (j = 1,2,3)$,
$f_j (v) = f_j (v + n \pi)$, we can slice the integration area $(0, \infty)$ to the
intercepts $(0, \pi), (\pi, 2 \pi), \dots (n \pi, (n+1) \pi) \dots$. Making
in each intercept the variable substitution $v \to v + n \pi$, we can rewrite:

\begin{eqnarray}
&&\int\limits_0^{\infty} \mathrm{d} v \, \exp (- \mathrm{i} \rho
v) \, f_j (v) = \sum\limits_{n=0}^\infty \exp (
- \mathrm{i} \rho n \pi) \, \int\limits_0^{\pi} \mathrm{d} v \, \exp
(- \mathrm{i} \rho v) \, f_j (v) 
\nonumber\\
&&= \frac{1}{1 - \exp (- \mathrm{i} \rho \pi)} \; A_j \,,
\label{eq:S(q)_int}
\end{eqnarray}
where
\begin{eqnarray}
A_j = \int\limits_0^{\pi} \mathrm{d} v \, \exp (- \mathrm{i} \rho
v) \, f_j (v) \,. \label{eq:S(q)_A_j}
\end{eqnarray}

It suffices to compute the integral $A_1$ because the other two integrals can
then be found with the help of the expressions:

\begin{eqnarray}
A_2 & = & \mathrm{i} \, \frac{\partial}{\partial \alpha} \, A_1
\,,\nonumber\\[2mm]
A_3 & = & - \frac{\mathrm{i}}{\alpha} \, \left( 1 - \mathrm{e}^{
- \mathrm{i} \rho \pi} \right) - \frac{\rho}{\alpha} \, A_1 \,.
\label{eq:S(q)_A_23}
\end{eqnarray}

One can easily verify the validity of the last formula substituting $A_3$ in the form:
\begin{equation}
A_3 = \frac{\mathrm{i}}{\alpha}\, \int\limits_0^{\pi}\mathrm{d}
v\,\exp (-\mathrm{i}\,\rho\,v)\, \frac{\mathrm{d}}{\mathrm{d}
v}\Bigl(\exp (-\mathrm{i}\,\alpha\,\mathrm{tan} v ) \Bigr)
\end{equation}
and computing the integral by parts.

It is not difficult to calculate the integral $A_1$ (see Appendix A for
details). Appling the formulas~(\ref{eq:S(q)_2}), (\ref{eq:S(q)_f_123}),
(\ref{eq:S(q)_A_j}), (\ref{eq:S(q)_A_23}) and the exact expression for $A_1$
(\ref{eq:S(q)_A_1}), we finally write the Fourier transform of the traslationally and
gauge invariant part of the electron propagator in the form:
\begin{eqnarray}
S (q) = \sum\limits_{n=0}^\infty \, \frac{\mathrm{i}}{q_{\|}^2 -
m_e^2 - 2 n \beta} \biggl\{\left[(q \gamma)_{\|} +
m_e\right] && \nonumber\\ \times \left[d_n (\alpha) -
\frac{\mathrm{i}}{2} \, (\gamma \varphi \gamma) \,
d_n^\prime(\alpha)\right] -(q \gamma)_{\perp} \, 2 n \, \frac{d_n
(\alpha)}{\alpha} \biggr\}\,,&&{}\label{eq:S(q)>}
\end{eqnarray}
\noindent where $\alpha = q_{\perp}^2/\beta$ and
\begin{equation}
d_n (\alpha) = (-1)^n \mathrm{e}^{-\alpha} [L_n (2 \alpha) - L_{n-1} (2 \alpha)]\,.
\label{eq:d_fun}
\end{equation}

%%%%%%%%%%%%%%%%%%%%%%%%%%%%%%%%%%%%%%%%%%%%%%%%%%%%%%%%%
\section{Propagators of the charged $W$ and $\Phi$-bosons expanded over the Landau levels}
\label{sec:PropagatorsLandau}
%%%%%%%%%%%%%%%%%%%%%%%%%%%%%%%%%%%%%%%%%%%%%%%%%%%%%%%%%

In a case when a magnetic field is sufficiently strong, $B \gtrsim B_e = m_e^2/e$,
it is convenient to use the electron propagator expanded over the Landau levels. The
procedure for obtaining this representation is described
in Ref.~\cite{Chodos:1990} (see also Refs.~\cite{Chyi:2000,Kuznetsov:2011_Okr}).
Note that there is a misprint in the expression for the propagator obtained
in Ref.~\cite{Chodos:1990}, namely the term on the second line in formula
(4.33) must contain the extra factor $(-\mathrm{i})$. This misprint was
corrected in later works, Ref.~\cite{Chyi:2000} (formulas (39) and (40)) and
also Ref.~\cite{Gusynin:1999} (formulas (13) and (14)) but without any comments.

The propagators of the $W$ and $\Phi$-bosons can be similarly expanded over the Landau
levels. As it was noted in the Introduction, in the conditions of the early Universe
magnetic fields could possibly exist with the scales of the critical magnetic
field value for the $W$-boson, $B_W = m_W^2/e \simeq 10^{24}$\,G. In this case,
a knowledge of the vector boson propagator in the form of expansion over the 
Landau levels may be useful.

The Fourier transform of the traslationally invariant part of the $W$-boson propagator 
in an arbitrary $\xi$-gauge is shown in Eq.~(\ref{eq:G(q)xi}). Let us
rewrite Eq.~(\ref{eq:G(q)xi}) in a more convenient way for further computations:
\begin{eqnarray}
&&G_{\mu \nu}(q)=
-\,\frac{1}{\beta}\int\limits_0^{\infty}\mathrm{d} v\,
\mathrm{e}^{-\mathrm{i}\,\rho\, v}\, \biggl[ (\tilde\varphi
\tilde\varphi)_{\mu \nu} \, \tilde{f}_1 (v) -(\varphi \varphi)_{\mu \nu} \, \tilde{f}_2 (v) - \varphi_{\mu \nu} \,
\tilde{f}_3 (v) \biggr] 
\nonumber\\
&&+ \frac{1}{\beta \,m_W^2}\int\limits_0^{\infty}\mathrm{d} v
\left( \mathrm{e}^{-\mathrm{i}\,\rho\, v} -
\mathrm{e}^{-\mathrm{i}\,\rho_\xi\, v}
\right) \biggl[ \biggl( q_\mu q_\nu + \mathrm{i} \,
\frac{\beta}{2} \, \varphi_{\mu \nu} \biggr)\, \tilde{f}_1 (v)  \label{eq:propGrs}\\
&&+ \biggl( (\varphi q)_\mu q_\nu + q_\mu (q \varphi)_\nu -
\mathrm{i} \, \frac{\beta}{2} \, (\varphi \varphi)_{\mu \nu}
\biggr) \, \tilde{f}_4 (v) + (\varphi
q)_\mu (q \varphi)_\nu \, \tilde{f}_5 (v) \biggr] \, , \nonumber
\end{eqnarray}
\noindent where we introduce the functions:
\begin{eqnarray}
\tilde{f}_1 (v) &=& \frac{1}{\cos v} \, \exp (- \mathrm{i} \, \alpha \,
\tan v) \,,
\nonumber\\[2mm]
\tilde{f}_2 (v) &=& \frac{\cos(2\,v)}{\cos v} \, \exp (- \mathrm{i} \,
\alpha \, \tan v) \,,
\nonumber\\[2mm]
\tilde{f}_3 (v) &=& \frac{\sin(2\,v)}{\cos v} \, \exp (- \mathrm{i} \,
\alpha \, \tan v) \,,
\\[2mm]\label{eq:S(q)_f_45678}
\tilde{f}_4 (v) &=& \frac{\tan v}{\cos v} \, \exp (- \mathrm{i} \, \alpha
\, \tan v) = \mathrm{i} \frac{\partial}{\partial \alpha} \, \tilde{f}_1 (v)
\,,
\nonumber\\[2mm]
\tilde{f}_5 (v) &=& \frac{\tan^2 v}{\cos v} \, \exp (- \mathrm{i} \, \alpha
\, \tan v) = - \frac{\partial^2}{\partial \alpha^2} \, \tilde{f}_1 (v)
\,,\nonumber
\end{eqnarray}
\noindent and also designate $\rho = ( m_W^2- q_{\|}^2 )/\beta$,
$\rho_\xi = ( \xi \, m_W^2- q_{\|}^2 )/\beta$, $\alpha =
q_{\perp}^2 /\beta$.

Taking into account that $\tilde{f}_j (v+\pi\,n) = (-1)^n\,\tilde{f}_j (v) \,
(j = 1,2,3,4,5)$, and slicing the integration area $(0, \infty)$ to the intercepts
$(0, \pi), (\pi, 2 \pi),\dots (n \pi, (n+1) \pi) \dots$ and making in each
intercept the variable substitution $v \to v + n \pi$, we can rewrite:

\begin{equation}
\int\limits_0^{\infty} \mathrm{d} v \, \exp (- \mathrm{i} \rho v)
\, \tilde{f}_j (v) = \frac{1}{1 + \exp (- \mathrm{i} \rho \pi)} \; \tilde{A}_j \,,
\label{eq:G(q)_int}
\end{equation}
\noindent where the integrals were introduced:
\begin{eqnarray}
\tilde{A}_j = \int\limits_0^{\pi} \mathrm{d} v \, \exp (- \mathrm{i} \rho
v) \, \tilde{f}_j (v), \, (j = 1,2,3,4,5)\,. \label{eq:G(q)_A_j}
\end{eqnarray}

In order to obtain the final expressions for~$\tilde{A}_1$, $\tilde{A}_2$,
$\tilde{A}_3$, it is necessary to use the relation 
$\sum_{n = 0}^{\infty} d_n(\alpha) = 1$ (assuming that $L_{-1}(x)\equiv 0$).
To present the results in a more compact way one should employ the notations:
\begin{eqnarray}
\ell_n(\alpha) &=& \frac{(n+1)\,d_{n+1}(\alpha) +
n\,d_n(\alpha)}{2 \,\alpha} = (-1)^n
\mathrm{e}^{-\alpha} L_n (2 \alpha)\,. \label{eq:ell_n_alpha}
\end{eqnarray}
\noindent The integrals take the following forms (see Appendix B for details):
\begin{eqnarray}
&\tilde{A}_1& = -2\,\mathrm{i} \left( 1 + \mathrm{e}^{- \mathrm{i} \rho
\pi} \right)
\sum\limits_{n = 0}^{\infty} \, \frac{\ell_{n-1}(\alpha)}{\rho+2\,n - 1}\,,\nonumber\\[5mm]
 &\tilde{A}_2& = -\mathrm{i} \left( 1 + \mathrm{e}^{
- \mathrm{i} \rho \pi} \right) \sum\limits_{n = 0}^{\infty}
\frac{\ell_n(\alpha) + \ell_{n-2}(\alpha)}{\rho+2\,n - 1}\,,\nonumber\\[5mm]
&\tilde{A}_3& = - \left( 1 + \mathrm{e}^{- \mathrm{i} \rho \pi} \right)
\sum\limits_{n = 0}^{\infty} \, \frac{\ell_n(\alpha) - \ell_{n-2}(\alpha)}{\rho+2\,n - 1}\,,\nonumber\\[5mm]
&\tilde{A}_4& = 2\left( 1 + \mathrm{e}^{- \mathrm{i} \rho \pi} \right)
\sum\limits_{n = 0}^{\infty} \, \frac{\ell'_{n-1}(\alpha)}{\rho+2\,n - 1}\,,\nonumber\\[5mm]
 &\tilde{A}_5& = 2\,\mathrm{i} \left( 1 + \mathrm{e}^{
- \mathrm{i} \rho \pi} \right) \sum\limits_{n = 0}^{\infty} \,
\frac{\ell''_{n-1}(\alpha)}{\rho+2\,n - 1}\,.\nonumber
\end{eqnarray}

Substituting the values of $\tilde{A}_{1}$, \dots $\tilde{A}_{5}$ into the expression for a propagator
(\ref{eq:propGrs}) we finally obtain: 
\begin{eqnarray}
&&G_{\mu \nu}(q) = \sum\limits_{n=0}^{\infty} \frac{
- \mathrm{i}}{q_{\|}^2 - m_W^2 - \beta\,(2\,n - 1)}
 \Biggl\{ 2\,(\tilde{\varphi}\,\tilde{\varphi})_{\mu \nu} \,
\ell_{n-1}(\alpha) 
\nonumber\\[2mm]
&&- \,(\varphi\,\varphi)_{\mu \nu} \,\Biggl(
\ell_n(\alpha) 
+ \ell_{n-2}(\alpha) \Biggr)+ \mathrm{i} \, \varphi_{\mu \nu}
\biggl(\ell_n(\alpha) - \ell_{n-2}(\alpha) \biggr)
\nonumber\\[2mm]
&& + \, \frac{\xi - 1}{q_{\|}^2 - \xi \, m_W^2 - \beta\,(2\,n - 1)}
\Biggl[ \biggl( 2\,q_\mu q_\nu + \mathrm{i} \, \beta \,
\varphi_{\mu \nu} \biggr)\, \ell_{n-1}(\alpha) 
\nonumber\\[2mm]
&&+ \, \mathrm{i}
\biggl( 2\,(\varphi q)_\mu q_\nu + 2\,q_\mu (q \varphi)_\nu
-\mathrm{i} \, \beta \, (\varphi \varphi)_{\mu \nu}\biggr) \, \ell'_{n-1}(\alpha) 
\nonumber\\[2mm]
&&- \, 2\,(\varphi q)_\mu (q \varphi)_\nu \, \ell''_{n-1}(\alpha)
 \Biggr]
 \Biggr\}\,.
\label{eq:G_over_n}
\end{eqnarray}

Note that unlike to the electron propagator, the contribution from the ground level
$n = 0$ into the propagator of the $W$-boson has a special feature. This contribution
has the gauge independent form:
\begin{equation}
G_{\mu \nu}^{(0)}(q) = \frac{- \mathrm{i}}{q_{\|}^2 - m_W^2 +
\beta} \, \mathrm{e}^{-q_{\perp}^2 /\beta} \, \left[
- (\varphi\,\varphi)_{\mu \nu} + \mathrm{i} \, \varphi_{\mu \nu}
\right]\,. \label{eq:G_over_0}
\end{equation}
\noindent It is essential that it contains a pole in the point $q_{\|}^2 = m_W^2 - \beta$.
So, if the magnetic field rises up to the critical value for the $W$-boson,
$B_W = m_W^2/e \simeq 10^{24}$\,G, then the so-called instability
arises of the perturbation theory for a $W$ boson vacuum, see Refs.
\cite{Skalozub:1987,Ambjorn:1989,Ambjorn:1990,MacDowell:1992}.

The propagator of the $\Phi$-boson in an arbitrary $\xi$-gauge, $D (q)$, can be obtained 
from  Eq.~(\ref{eq:G_over_n}), similarly to the case of the weak field approximation, in the form:

\begin{eqnarray}
D (q) = \sum\limits_{n=0}^{\infty} \frac{2\,\mathrm{i} \,
\ell_{n-1}(\alpha)}{q_{\|}^2 - \xi \, m_W^2 - \beta\,(2\,n - 1)}
\,. \label{eq:D(q)_over_n}
\end{eqnarray}
\noindent Note that the summation over $n$ in the formula~(\ref{eq:D(q)_over_n})
starts formally from $n=0$, but actually from $n=1$, because $\ell_{-1}(\alpha) = 0$ by definition. 
This means that the propagator of the $\Phi$ boson, as one could expect, does not contain a pole at 
the point $q_{\|}^2 = \xi \, m_W^2 - \beta$.

%%%%%%%%%%%%%%%%%%%%%%%%%%%%%%%%%%%%%%%%%%%%%%%%%%%%%%%%%
\section{Conclusion}
\label{sec:Conclusion}
%%%%%%%%%%%%%%%%%%%%%%%%%%%%%%%%%%%%%%%%%%%%%%%%%%%%%%%%%

Different representations of the propagators of charged particles in a constant
uniform magnetic field have been analyzed. The expressions for the propagators
of the vector $W$-boson and the scalar $\Phi$-boson of the Standard Model in an arbitrary
$\xi$-gauge expanded over the Landau levels, Eqs.~(\ref{eq:G_over_n}) 
and~(\ref{eq:D(q)_over_n}), were derived for the first time.

%%%%%%%%%%%%%%%%%%%%%%%%%%%%%%%%%%%%%%%%%%%%%%%%%%%%%%%%%
\section*{Acknowledgments}
%%%%%%%%%%%%%%%%%%%%%%%%%%%%%%%%%%%%%%%%%%%%%%%%%%%%%%%%%

We wish to thank V.A. Rubakov and all the participants of the seminar at the Theoretical Physics Department of the Institute for Nuclear Research of the Russian Academy of Sciences for a stimulating discussion.

This study was supported by the Russian Foundation for Basic Research (Project No. \mbox{14-02-00233-a}). 
The work of A.V.K. was performed with the support by the Project No.~92 within the base part of the State Assignment 
for the Yaroslavl University Scientific Research.

%%%%%%%%%%%%%%%%%%%%%%%%%%%%%%%%%%%%%%%%%%%%%%%%%%%%%%%%%
\section*{Note Added in Proof}
%%%%%%%%%%%%%%%%%%%%%%%%%%%%%%%%%%%%%%%%%%%%%%%%%%%%%%%%%

After the paper was completed, we have been aware of the paper \cite{Ayala:2005}
where the formula for the propagator of a charged scalar particle expanded over Landau levels
was obtained. Our expression for this propagator, taken in the Feynman gauge, coincides with
the result of Ref. \cite{Ayala:2005}, to the notations. We thank A. Ayala for informing us on their paper.

%%%%%%%%%%%%%%%%%%%%%%%%%%%%%%%%%%%%%%%%%%%%%%%%%%%%%%%%%
\appendix

%%%%%%%%%%%%%%%%%%%%%%%%%%%%%%%%%%%%%%%%%%%%%%%%%%%%%%%%%
\section{Calculation of the integral $A_{1}$}
%%%%%%%%%%%%%%%%%%%%%%%%%%%%%%%%%%%%%%%%%%%%%%%%%%%%%%%%%

To compute $A_1$ let us rewrite $f_1 (v)$ in the following form:

\begin{eqnarray}
f_1 (v) = \exp (- \mathrm{i} \, \alpha \, \tan v) = \exp
\left(\alpha \, \frac{- \mathrm{e}^{-2 \mathrm{i} v} +1 }{
- \mathrm{e}^{-2 \mathrm{i} v} - 1 } \right) \,.
\label{eq:S(q)_A_j1}
\end{eqnarray}

The right-hand side of this equation can be expressed through the Laguerre polynomials:
\begin{equation}
L_n (x) = \frac{1}{n!} \, \mathrm{e}^{x} \,
\frac{\mathrm{d}^n}{\mathrm{d} x^n} \left( x^n \,
\mathrm{e}^{-\,x} \right) . \label{eq:ChebLag_}
\end{equation}
The generating function for the Laguerre polynomials is determined by

\begin{equation}
\frac{1}{1-t} \, \exp \left( - \frac{x \, t}{1-t} \right) =
\sum\limits_{n=0}^\infty L_n (x) \, t^n \label{eq:ChebLag_gen_fun}
\end{equation}
for $|t| < 1$, so
\begin{equation}
\exp \left( - \frac{x \, t}{1-t} \right) =
\sum\limits_{n=0}^\infty \left[ L_n (x) - L_{n-1} (x) \right] \,
t^n \,, \label{eq:ChebLag_gen_fun2}
\end{equation}
where we assume $L_{-1} (x) \equiv 0$. Denoting in the right part of the
formula~(\ref{eq:S(q)_A_j1}): $- \mathrm{e}^{-2 \mathrm{i} v} = t$,
$2 \alpha = x$, and using the identity
\begin{equation}
\exp \left( \frac{x}{2} \, \frac{t + 1}{t - 1} \right) = \exp
\left( - \frac{x \, t}{1-t} \right) \, \exp \left( - \frac{x}{2}
\right) \,, \label{eq:ChebLag_gen_fun3}
\end{equation}
\noindent let us transform $A_1$ to the form
\begin{eqnarray}
&& A_1 = \int\limits_0^{\pi} \mathrm{d} v \, \mathrm{e}^{- \alpha}
\, \sum\limits_{n=0}^\infty \bigg[ L_n (2 \alpha) -
L_{n-1} (2 \alpha) \bigg] (- 1)^n\,\exp (-2 \mathrm{i} n v) \,
\exp (- \mathrm{i} \rho v) 
\nonumber\\
&& = \mathrm{e}^{- \alpha} \, \sum\limits_{n=0}^\infty \, (- 1)^n
\left[ L_n (2 \alpha) - L_{n-1} (2 \alpha)
\right]\int\limits_0^{\pi} \mathrm{d} v \, \exp [- \mathrm{i}
(\rho + 2 n) v] 
\nonumber\\
&&= - \mathrm{i} \, \mathrm{e}^{- \alpha} \, \left( 1 -
\mathrm{e}^{- \mathrm{i} \rho \pi} \right)
\sum\limits_{n=0}^\infty \, \frac{(- 1)^n}{\rho + 2 n} \bigg[ L_n
(2 \alpha) - L_{n-1} (2 \alpha) \bigg] \,. \label{eq:S(q)_A_1}
\end{eqnarray}

%%%%%%%%%%%%%%%%%%%%%%%%%%%%%%%%%%%%%%%%%%%%%%%%%%%%%%%%%
\section{Calculation of the functions $\tilde{A}_{j}$}
%%%%%%%%%%%%%%%%%%%%%%%%%%%%%%%%%%%%%%%%%%%%%%%%%%%%%%%%%

In order to calculate the functions $\tilde{A}_{j}$, it is worthwhile to introduce the auxiliary integrals:
\begin{eqnarray}
\mathrm{C}(\alpha) &=&
\int\limits_0^{\pi} \mathrm{d} v\, \exp (-\mathrm{i}\,\rho\,v)\, \exp (- \mathrm{i} \, \alpha \, \mathrm{\tan} v )\, \cos v,\nonumber\\[3mm]
\mathrm{S}(\alpha) &=&
\int\limits_0^{\pi} \mathrm{d} v\, \exp (-\mathrm{i}\,\rho\,v)\, \exp (- \mathrm{i} \, \alpha \, \mathrm{tan} v )\, \sin v,\nonumber\\[3mm]
\mathrm{E^{(\pm)}} (\alpha) &=& \mathrm{C}(\alpha) \pm \mathrm{i}
\, \mathrm{S}(\alpha) =
\int\limits_0^{\pi} \mathrm{d} v\, \exp\bigl[-\mathrm{i}
\,(\rho \mp 1)\,v\bigr] \exp (- \mathrm{i} \, \alpha \,
\mathrm{\tan} v)\,.\nonumber
\end{eqnarray}
\noindent $\tilde{A}_1$ can be represented in the form
\begin{equation}
\tilde{A}_1 = \frac{\mathrm{i}}{\alpha}\, \int\limits_0^{\pi}\mathrm{d}
v\,\exp (-\mathrm{i}\,\rho\,v)\, \cos v \,
\frac{\mathrm{d}}{\mathrm{d} v}\Bigl(\exp
(-\mathrm{i}\,\alpha\,\mathrm{\tan} v ) \Bigr)
\end{equation}
\noindent and further, integrating by parts we obtain:
\begin{equation}
\tilde{A}_1 = \frac{\mathrm{i}}{\alpha}\, \biggl[- 1 - \exp (- \mathrm{i}
\rho \pi) + \mathrm{i}
\,\rho\,\mathrm{C}(\alpha)+\mathrm{S}(\alpha)\biggr]\,.
\end{equation}
The integrals $\tilde{A}_2$ and $\tilde{A}_3$ can be expressed through $\tilde{A}_1$,
$\mathrm{C}(\alpha)$ and $\mathrm{S}(\alpha)$:
\begin{eqnarray}
&& \tilde{A}_2 = 2\,\mathrm{C}(\alpha) - \tilde{A}_1,\\[5mm]
&& \tilde{A}_3 = 2\,\mathrm{S}(\alpha)\,.
\end{eqnarray}
\noindent In order to calculate $\mathrm{C}(\alpha)$ and
$\mathrm{S}(\alpha)$, let us compute $\mathrm{E^{(\pm)}}(\alpha)$ and use the expressions:

\begin{eqnarray}\label{eq:C.int}
&& \mathrm{C}(\alpha) = \frac{1}{2}\,\left[E^{(+)}(\alpha) + E^{(-)}(\alpha)\right],\\[5mm]
\label{eq:S.int} && \mathrm{S}(\alpha) =
\frac{1}{2\,\mathrm{i}}\,\left[E^{(+)}(\alpha) - E^{(-)}(\alpha)\right] \,.
\end{eqnarray}
\noindent The integral $\mathrm{E^{(\pm)}}(\alpha)$ equals to

\begin{equation}
\mathrm{E^{(\pm)}}(\alpha) = -\mathrm{i} \left[ 1 + \exp (
-\mathrm{i} \rho \pi) \right]
\sum\limits_{n=0}^{\infty}\frac{d_n(\alpha)}{\rho+2\,n \mp 1}\,.
\end{equation}
\noindent The functions $d_n(\alpha)$ are determined by the equation \eqref{eq:d_fun}.

\noindent The integrals $\mathrm{C}(\alpha)$ and $\mathrm{S}(\alpha)$
can be expressed as
\begin{equation}
 \mathrm{C}(\alpha) = -\,\frac{\mathrm{i}}{2} \left[ 1 + \exp (- \mathrm{i} \rho \pi) \right] \sum\limits_{n=0}^{\infty}\frac{d_n(\alpha) + d_{n-1}(\alpha)}{\rho+2\,n - 1}\,,\\[5mm]
\end{equation}\vspace{-4pt}
\begin{equation}
 \mathrm{S}(\alpha) = -\,\frac{1}{2} \left[ 1 + \exp (- \mathrm{i} \rho \pi) \right] \sum\limits_{n=0}^{\infty}\frac{d_n(\alpha) - d_{n-1}(\alpha)}{\rho+2\,n - 1}\,.
\end{equation}

%%%%%%%%%%%%%%%%%%%%%%%%%%%%%%%%%%%%%%%%%%%%%%%%%%%%%%%%%

%%%%%%%%%%%%%%%%%%%%%%%%%%%%%%%%%%%%%%%%%%%%%%%%%%%%%%%%%

\begin{thebibliography}{00} %for 2 digits
%%%%%%%%%%%%%%%%%%%%%%%%%%%%%%%%%%%%%%%%%%%%%%%%%%%%%%%%%

\bibitem{Duncan:1992}
 R.\,C. Duncan and C. Thompson,
 \textit{Astrophys. J.} \textbf{392}, L9 (1992).

\bibitem{McGillOnline} 
 McGill Pulsar Group SGR/AXP Online Catalog,
\newline  \textrm{http://www.physics.mcgill.ca/\~}\!\! \textrm{pulsar/magnetar/main.html}.

\bibitem{Kunze:2013}
K.\,E. Kunze, 
\textit{Plasma Phys. Control. Fusion} \textbf{55}, 124026 (2013). % , arXiv:1307.2153 [astro-ph.CO].

\bibitem{Vachaspati:1991}
T. Vachaspati,
 \textit{Phys. Lett. B} \textbf{265}, 258 (1991).
%
\bibitem{Ambjorn:1992}
J. Ambj{\o}rn and P. Olesen,
 ``Electroweak magnetism, $W$-codensation and anti-screening'',
 in: \textit{Proc. of 4th Hellenic School on Elementary
 Particle Physics}, Corfu, 1992, arXiv: hep-ph/9304220.
%
\bibitem{Grasso:2001}
D. Grasso and H.\,R. Rubinstein,
% Magnetic fields in the early Universe,
\textit{Phys. Rep.} \textbf{348}, 163 (2001). % arXiv:astro-ph/0009061

\bibitem{Giovannini:2004}
M. Giovannini, 
%The Magnetized Universe,
\textit{Int. J. Mod. Phys. D} \textbf{13}, 391 (2004). % arXiv:astro-ph/0312614.

\bibitem{Kandus:2011} 
   A. Kandus, K.\,E. Kunze and C.\,G. Tsagas, %Primordial magnetogenesis
   \textit{Phys. Rep.} \textbf{505}, 1 (2011).  

\bibitem{Skokov:2009} 
V. Skokov, A. Illarionov and V. Toneev, 
\textit{Int.~J.~Mod.~Phys.~A} \textbf{24}, 5925 (2009). % arXiv:0907.1396 [nucl-th]

\bibitem{Ritus:1979}
V.\,I.~Ritus, Quantum effects of the interaction of elementary particles with an intense electromagnetic field (in Russian), in \textit{Quantum Electrodynamics of Phenomena in an Intense Field}, Proc. P.\,N.~Lebedev Physical Institute, vol.~111 (Nauka, Moscow, 1979), pp.~5--151.
%
\bibitem{Itzykson:1980}
C.~Itzykson and J.-B.~Zuber, \textit{Quantum Field Theory} (McGraw-Hill, New~York, 1985).
%
\bibitem{Papanian:1986}
V.\,O.~Papanian and V.\,I.~Ritus, Three-photon interaction in an intense field, in \textit{Issues in Intense-Field Quantum Electrodynamics}, ed. by V.\,L.~Ginzburg (Nova Science Publishers, New~York, 1989), pp.~153--179.
%
\bibitem{Shabad:1988}
A.\,E.~Shabad, in \textit{Polarization of the Vacuum and a~Quantum Relativistic Gas in an External Field}, ed. by V.\,L.~Ginzburg (Nova Science Publishers, New~York, 1992).
%
\bibitem{Ternov:1989}
I.\,M.~Ternov, V.\,Ch.~Zhukovskii and A.\,V.~Borisov, \textit{Quantum Processes in Strong External Field} (Moscow State Univ., Moscow, 1989), in Russian.
%
\bibitem{Kuznetsov:2003}
A.\,V.~Kuznetsov and N.\,V.~Mikheev, \textit{Electroweak Processes in External Electromagnetic Fields} (Springer-Verlag, New~York, 2003).

\bibitem{Schwinger:1951}
J. Schwinger,
 \textit{Phys. Rev.} \textbf{82}, 664 (1951).

\bibitem{Fock:1937}
V.\,A. Fock,
 \textit{Phyzik. Z. Sowjetunion} \textbf{12}, 404 (1937).

\bibitem{Loskutov:1976}
Yu.\,M. Loskutov and V.\,V. Skobelev,
 \textit{Phys.~Lett.~A} \textbf{56}, 151 (1976).

\bibitem{Chodos:1990}
   A. Chodos, K. Everding and D.\,A. Owen,
   \textit{Phys. Rev. D} \textbf{42}, 2881 (1990).

\bibitem{Chyi:2000} 
   T.-K. Chyi, C.-W. Hwang, W.F. Kao, G.-L. Lin, K.-W. Ng and J.-J. Tseng,
   \textit{Phys. Rev. D} \textbf{62}, 105014 (2000). 

\bibitem{Kuznetsov:2011_Okr} 
   A.\,V. Kuznetsov and A.\,A. Okrugin,
   \textit{Int. J. Mod. Phys. A} \textbf{26}, 2725 (2011). 

\bibitem{Elizalde:2002}
   E. Elizalde, E.\,J. Ferrer and V. de la Incera, 
   \textit{Ann. of Phys.} \textbf{295}, 33 (2002).

\bibitem{Elizalde:2004}
   E. Elizalde, E.\,J. Ferrer and V. de la Incera, 
   \textit{Phys. Rev. D} \textbf{70}, 043012 (2004).

\bibitem{Kuznetsov:2006}
   A.\,V. Kuznetsov, N.\,V. Mikheev, G.\,G. Raffelt and L.\,A. Vassilevskaya, 
   \textit{Phys. Rev. D} \textbf{73}, 023001 (2006).

\bibitem{Erdas:2003} 
   A. Erdas and M. Lissia,
   \textit{Phys. Rev. D} \textbf{67}, 033001 (2003).

\bibitem{Bhattacharya:2009}
   K. Bhattacharya and S. Sahu,
   \textit{Eur. Phys. J. C} \textbf{62}, 481 (2009).

\bibitem{Kuznetsov:2010_PLB}
   A.\,V. Kuznetsov, N.\,V. Mikheev and A.\,V. Serghienko, 
   \textit{Phys. Lett. B} \textbf{690}, 386 (2010).

\bibitem{Erdas:2000}
   A. Erdas and C. Isola, 
   \textit{Phys. Lett. B} \textbf{494}, 262 (2000).

\bibitem{Erdas:1990}
   A. Erdas and G. Feldman, 
   \textit{Nucl. Phys. B} \textbf{343}, 597 (1990).

\bibitem{Gusynin:1999}
   V.\,P. Gusynin and A.\,V. Smilga,
   \textit{Phys. Lett. B} \textbf{450}, 267 (1999).
   
\bibitem{Skalozub:1987}
   V.\,V. Skalozub, 
   \textit{Yad. Fiz.} \textbf{45}, 1708 (1987) 
   [\textit{Sov. J. Nucl. Phys.} \textbf{45}, 1058 (1987)].

\bibitem{Ambjorn:1989}
J. Ambj{\o}rn and P. Olesen,
   \textit{Nucl. Phys. B} \textbf{315}, 606 (1989).

\bibitem{Ambjorn:1990}
J. Ambj{\o}rn and P. Olesen,
   \textit{Int. J. Mod. Phys. A} \textbf{5}, 4525 (1990).

\bibitem{MacDowell:1992}
S.\,W. MacDowell and O. T\"ornkvist, 
   \textit{Phys. Rev. D} \textbf{45}, 3833 (1992).

\bibitem{Ayala:2005}
A. Ayala, A. S{\'a}nchez, G. Piccinelli and S. Sahu,
  \textit{Phys. Rev. D} \textbf{71}, 023004 (2005).
\end{thebibliography}
\end{document}